\documentclass[12pt]{article}   
\usepackage{epsfig}
\def\Journal#1#2#3#4{{#1} {\bf #2}, #3 (#4)}

\def\PLB{Phys. Lett. B}
\def\PRL{Phys. Rev. Lett.}
\def\PRD{Phys. Rev. D}
\def\ZPC{Z. Phys. C}
\def\EPJ{Eur. Phys. J. C}

\def\ra{\rightarrow}

\newcommand{\SC}{\langle{\cal O}_8^{\psi^{\prime}}(^1S_0)\rangle}
\newcommand{\SD}{\langle{\cal O}_8^{\psi^{\prime}}(^3S_1)\rangle}
\newcommand{\SDsing}{\langle{\cal O}_1^{\psi^{\prime}}(^3S_1)\rangle}
\newcommand{\PS}{\langle{\cal O}_8^{\psi^{\prime}}(^3P_0)\rangle}
\newcommand{\PJ}{\langle{\cal O}_8^{\psi^{\prime}}(^3P_J)\rangle}

\begin{document}
\begin{flushright}
KOBE--FHD--01--04\\
May~~~2001
\end{flushright}

\begin{center}
\baselineskip=1cm
{\Large \bf Electroproduction of $\psi^{\prime}$ \\
and polarized gluon distribution in a proton} 

\baselineskip=0.6cm
\vspace{1.5cm}
Kazutaka ~SUDOH \\
\vspace{0.2cm}
{\em Graduate School of Science and Technology, \\ 
Kobe University, Nada, Kobe 657--8501, Japan} \\
E-mail: sudou@radix.h.kobe-u.ac.jp

\vspace{0.8cm}
Han--Wen ~HUANG$^{a,}$\footnote[2]
{Postdoctoral research fellow (No.P99221) of the Japan Society for 
the Promotion of Science (JSPS).} 
~and ~Toshiyuki ~MORII$^{b}$ \\
\vspace{0.2cm}
{\em Faculty of Human Development,  
Kobe University, Nada, Kobe 657--8501, Japan} \\
$^{a}$E-mail: huanghw@radix.h.kobe-u.ac.jp \\
$^{b}$E-mail: morii@kobe-u.ac.jp
\end{center}

\vspace{1.0cm} 
\noindent
\begin{center}
{\bf Abstract}
\end{center}

In order to get information about the polarized gluon 
distribution in a proton, we studied the electroproduction of 
$\psi^{\prime}$ in polarized electron and polarized proton collisions 
in the framework of the NRQCD factorization approach. 
The value of the cross section $d\Delta\sigma / dp_T^2$ for color--octet 
mechanism is about 5 times larger than that for color--singlet one, 
and it might be another test of the color--octet model if the polarized 
gluon distribution $\Delta g(x)$ is well known. 
Furthermore, we found that this reaction is quite effective for testing 
the model of gluon polarization by measuring the double spin asymmetry 
$A_{LL}$ for the initial electron and proton. 
Though the shape of $p_T^2$ distribution of $A_{LL}$ is quite similar 
for the production mechanism with color--octet and color--singlet, 
we can see a big difference in $A_{LL}$ among various models of 
the polarized gluon distribution function $\Delta g(x)$. 

\clearpage

The proton spin puzzle \cite{spin} has been one of the most 
challenging topics in high energy hadron physics. 
As is well known, the proton spin is composed of the spin and angular 
momenta of quarks and gluons which constitute a proton. 
To understand the spin structure of a proton, it is necessary to 
study the behavior of the parton distribution functions which plays 
an important role in describing the spin structure of a proton. 
So far, a number of good parameterization models of the spin--independent 
parton distributions have been already obtained. 
In these years, we have also a large number of experimental data 
on the spin--dependent structure functions $g_1 (x, Q^2)$ \cite{g_1} 
which lead to extensive study on the spin--dependent parton distributions. 
However, knowledge of the polarized gluon distribution is still poor, 
though many processes have been suggested so far for extracting 
information about it. 

On the other hand, since the advent of surprisingly big cross sections 
of prompt $J/\psi$ and $\psi^{\prime}$ hadroproduction at large $p_T$ 
regions observed by the CDF collaboration \cite{DATA1}, production 
mechanism of heavy quarkonium has been one of the most challenging 
topics in current particle physics with QCD. 
The CDF results were larger than the prediction by the conventional 
color--singlet model by more than one order of magnitude. 
$J/\psi$ photoproduction data at $ep$ collider HERA experiment are 
also at variance with the prediction by the color--singlet 
model \cite{DATA2}. 
In order to solve these serious discrepancies between the experimental 
data and the theoretical prediction by the color--singlet 
model \cite{Braaten95}, the NRQCD factorization formalism has been 
recently proposed \cite{Bodwin95} as one of the most promising candidates. 
The NRQCD factorization approach separates the effects of short distance 
scales that are comparable to or smaller than the inverse of heavy quark 
mass from the ones of longer distance scales that cause 
the hadronization of heavy quark pair. 
A heavy quark and its antiquark pair is allowed to 
be produced not only in color--singlet but also in color--octet 
intermediate state at a short distance of the process, and 
subsequently the color--octet pair hadronizes into a final 
color--singlet quarkonium via emission or absorption of dynamical gluons, 
which is called the color--octet mechanism. 
A short distance coefficient can be computed using perturbative QCD, 
and a long distance parameter is described by nonperturbative NRQCD 
matrix elements whose values should be determined from experiments or 
lattice gauge theory. 
Unfortunately, the present uncertainties of the NRQCD matrix elements are 
not so small and thus the discussion on whether the NRQCD works well 
or not is quite controversial. 
There have been several discussions that to test the color--octet 
mechanism, it is important to study the heavy quarkonium production in 
various polarized reactions \cite{Sudoh00}. 

\begin{figure}[t]
\begin{center}
\hspace*{0.1cm}
  \epsfxsize=7cm
  \epsfbox{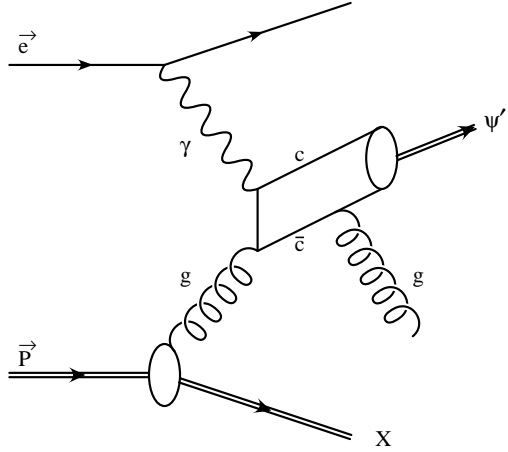}
\end{center}
\vspace{-0.4cm}
\caption[junk]{Feynman diagram for large--$p_T$ $\psi^{\prime}$
leptoproduction at the leading order.
Incident electron and proton are polarized. 
\label{fig:feyndiagram_ep1}}
\end{figure}

In this work, to extract information about the polarized gluon 
density in a proton, we study the polarized electroproduction of 
$\psi^{\prime}$ at large--$p_T$ regions by taking account of 
color--octet contribution, 
\begin{equation}
\vec{e} + \vec{p} \ra \psi^{\prime} + X , 
\label{eqn:leptopro}
\end{equation}
which could be observed in the forthcoming $e$--RHIC or TESLA--$\vec{N}$ 
experiments, where incident electron and proton are longitudinally polarized. 
Typical Feynman diagram of this process is illustrated in 
Fig. \ref{fig:feyndiagram_ep1}. 
Since the process is dominated by photon-gluon fusion, the cross 
section must be sensitive to the gluon density in a proton and thus 
one can obtain good information about the polarized gluon 
distribution function. 
%
\noindent

An interesting observable is the double spin asymmetry $A_{LL}$ 
which is defined by 
\begin{equation}
A_{LL} (\beta) = \frac{d\Delta\sigma / d\beta}{d\sigma / d\beta} ,
\end{equation}
where $\beta$ is some kinematical variables. 
$d\Delta\sigma$ represents spin--dependent cross section and is defined by 
\begin{equation}
d\Delta\sigma=\frac{1}{4}\left[d\sigma_{++} - d\sigma_{+-} 
+ d\sigma_{--} - d\sigma_{-+}\right] , 
\end{equation}
where $d\sigma_{+-}$, for instance, denotes the cross section with 
positive electron helicity and negative proton helicity. 
The asymmetry $A_{LL}$ is generally more sensitive to the 
mechanism of hard processes than the cross section itself. 
This is because, since the asymmetry is normalized, the input 
parameters such as the quark mass, strong coupling constant, etc. which 
are not related to dynamics, are dropped out from the numerator 
and the denominator. 

The cross sections of large--$p_T$ $J/\psi$ production in $\gamma p$ 
collisions were calculated for the unpolarized 
case \cite{Beneke98, Ko96} 
and the polarized case \cite{Yuan00, Japaridze99}. 
Here we use the Weizs\"acker--Williams approximation
\cite{Weizsacker34} to evaluate the cross sections for the 
process (\ref{eqn:leptopro}). 
In this scheme we can write the electroproduction cross section in terms 
of the photoproduction cross section convoluted by the photon flux factor 
in the electron, 
\begin{equation}
d\Delta\sigma(ep \ra \psi^{\prime} + X)
=\int dy \Delta f_{\gamma/e}(y) 
d\Delta\sigma(\gamma p \ra \psi^{\prime} + X) , 
\label{eqn:ep2}
\end{equation}
where $y$ is the energy fraction of electron carried by photon, 
defined by 
$y=p_{{}_{P}}\cdot q_{\gamma}/p_{{}_{P}}\cdot p_{e}$ with 
$p_{{}_{P}}$, $p_{e}$, and $q_{\gamma}$ being the momentum 
of proton, electron, and photon, respectively. 
In this calculation, we have neglected the contribution from the resolved 
photon process, which can be safely eliminated by introducing 
appropriate kinematical cuts \cite{Beneke98}. 
$\Delta f_{\gamma/e}(y)$ is the polarized photon distribution function 
in the electron, which is taken as 
\cite{deFlorian99} 
\begin{equation}
\Delta f_{\gamma/e}(y)=\frac{\alpha_{em}}{2\pi}
\left[
\frac{1-(1-y)^2}{y}\log\frac{Q_{max}^{2}(1-y)}{m_e^{2}y^2}
+2m_e^{2}y^2\left(\frac{1}{Q_{max}^2}-\frac{1-y}{m_e^{2}y^2} \right)
\right] , 
\end{equation}
where $m_e$ is the electron mass. 
$Q^2_{max}$ is the maximum value of $Q^2$ for photoproduction 
processes and we adopt $Q^2_{max} =1$ GeV as in Ref \cite{deFlorian99}. 

The process is dominated by photon--gluon fusion. 
We neglected here the contribution from the photon--quark fusion 
process, since its contribution to the cross sections is only a few 
percent \cite{Yuan00}. 
Photoproduction cross section in the integrand of Eq. (\ref{eqn:ep2}) 
is calculated by using the subprocess cross section as 
\begin{equation}
d\Delta\sigma(\gamma p \ra \psi^{\prime} + X)
=\int dx\Delta g(x,Q^2)d\Delta\hat{\sigma}(\gamma g \ra \psi^{\prime} + X) ,
\end{equation}
where $\Delta g(x,Q^2)$ is the polarized gluon distribution function 
in a proton. 

In the NRQCD factorization formalism, the subprocess cross section can 
be factorized into the short and long distance factors as 
\begin{equation}
d\Delta\hat{\sigma}(\gamma g \ra \psi^{\prime} + X)
=\sum_n \Delta C_{n}
\langle {\cal O}_{1, 8}^{\psi^{\prime}}(^{2S+1}L_{J})\rangle .
\end{equation}
The label $n$ represents color and angular momentum configuration of 
intermediate $c\bar{c}$ pair. 
$\Delta C_{n}$ is a short distance coefficient for the polarized 
process and can be calculated as perturbation series with a QCD 
coupling constant $\alpha_s$. 
$\langle {\cal O}_{1, 8}^{\psi^{\prime}}(^{2S+1}L_{J})\rangle$ is 
the vacuum expectation value of the NRQCD matrix element, whose relative 
importance is determined by the NRQCD velocity scaling rule. 
This long distance nonperturbative parameter represents the probability 
for $c\bar{c}$ pair being in the $n$ configuration which evolves into 
physical state $\psi^{\prime}$, and can be extracted from experiments 
at present. 
%
\noindent
\begin{figure}[t]
\begin{center}
\hspace*{0.1cm}
  \epsfxsize=5cm
  \epsfbox{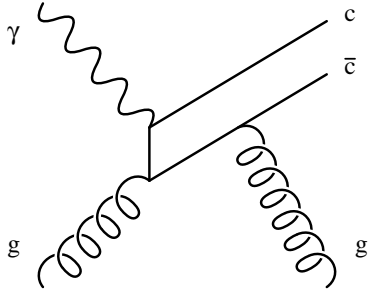}
\end{center}
\vspace{-0.4cm}
\caption[junk]{Feynman diagram for the leading order subprocess 
$\gamma + g \ra (c\bar{c}) +g$ 
at large--$p_T$ regions. 
\label{fig:feyndiagram_ep2}}
\end{figure}

At large--$p_T$ regions, the process is dominated by the subprocess 
$\gamma^* + g \ra (c\bar{c}) + g$ as illustrated in 
Fig. \ref{fig:feyndiagram_ep2}. 
At the leading order, we can expect the contributions from 
the following 4 processes: 
\begin{equation}
\gamma + g \ra c\bar{c} (^3S_{1}, \underline{1}) + g ,
\end{equation}
\begin{equation}
\gamma + g \ra c\bar{c} (^1S_{0}, \underline{8}) + g ,
\end{equation}
\begin{equation}
\gamma + g \ra c\bar{c} (^3S_{1}, \underline{8}) + g ,
\end{equation}
\begin{equation}
\gamma + g \ra c\bar{c} (^3P_{J}, \underline{8}) + g ,
\end{equation}
each of which has an intermediate $c\bar{c}$ pair with definite angular momentum 
and color. 
The first process is the conventional color--singlet process. 
The other three processes are color--octet processes induced by the 
NRQCD factorization formalism. 

For numerical values of the NRQCD matrix elements, we used the results 
of recent analysis by \cite{Beneke96} and \cite{Braaten99}: 
\begin{equation}
\SDsing =6.70\times 10^{-1}~~~{\rm GeV}^3 , 
\label{eqn:singlet}
\end{equation}
\begin{equation}
\SC =0.75\times 10^{-2}~~~{\rm GeV}^3 , 
\end{equation}
\begin{equation}
\SD =0.37\times 10^{-2}~~~{\rm GeV}^3 , 
\end{equation}
\begin{equation}
\PS /m_c^2 =0.01\times 10^{-2}~~~{\rm GeV}^3 , 
\label{eqn:P-wave}
\end{equation}
\begin{equation}
\PJ =(2J+1)\PS . 
\end{equation}
The color--singlet matrix element given in Eq. (\ref{eqn:singlet}) 
is related to the radial wave function at the origin, 
whose value is extracted from potential model calculations or 
directly from experiments. 
In fact, its value can be determined from the leptonic decay width of 
$\psi^{\prime}$ with good accuracy. 
The values of the color--octet matrix elements were taken from the 
analysis on charmonium hadroproduction data. 
However, their values have not well been determined at present and 
various set of parameter values have been proposed \cite{Braaten99}. 
The main source of ambiguity of those matrix elements comes from the 
higher order correction and uncertainties of unpolarized parton 
distribution functions. 
Especially, the value of $P$--wave color--octet matrix element 
$\PS$ for the $\psi^{\prime}$ production becomes negative in some 
regions, which is unphysical. 
For example, combining the data of Ref \cite{Beneke96} with the Type I 
set of Ref \cite{Braaten99} and the data of Ref \cite{Beneke96} with 
the Type II set of Ref \cite{Braaten99}, we obtain the value of $\PS$ as 
\begin{eqnarray}
\PS /m_c^2 &=& -0.22 \pm 0.13\times 10^{-2}~~~{\rm GeV}^3 
~~~~{\rm for~~Type~~I}, \\
\PS /m_c^2 &=& -0.08 \pm 0.11\times 10^{-2}~~~{\rm GeV}^3 
~~~~{\rm for~~Type~~II}.
\end{eqnarray}
As shown above, though the value of $\PS$ for Type I set is negative 
and unphysical, there can be a possibility of positive value of $\PS$ 
for Type II set by taking its observed error into consideration. 
Here we have adopted Eq. (\ref{eqn:P-wave}) as a possible parameter for 
Type II set. 

\begin{figure}[t]
\begin{center}
    \epsfxsize=7cm
    \epsfbox{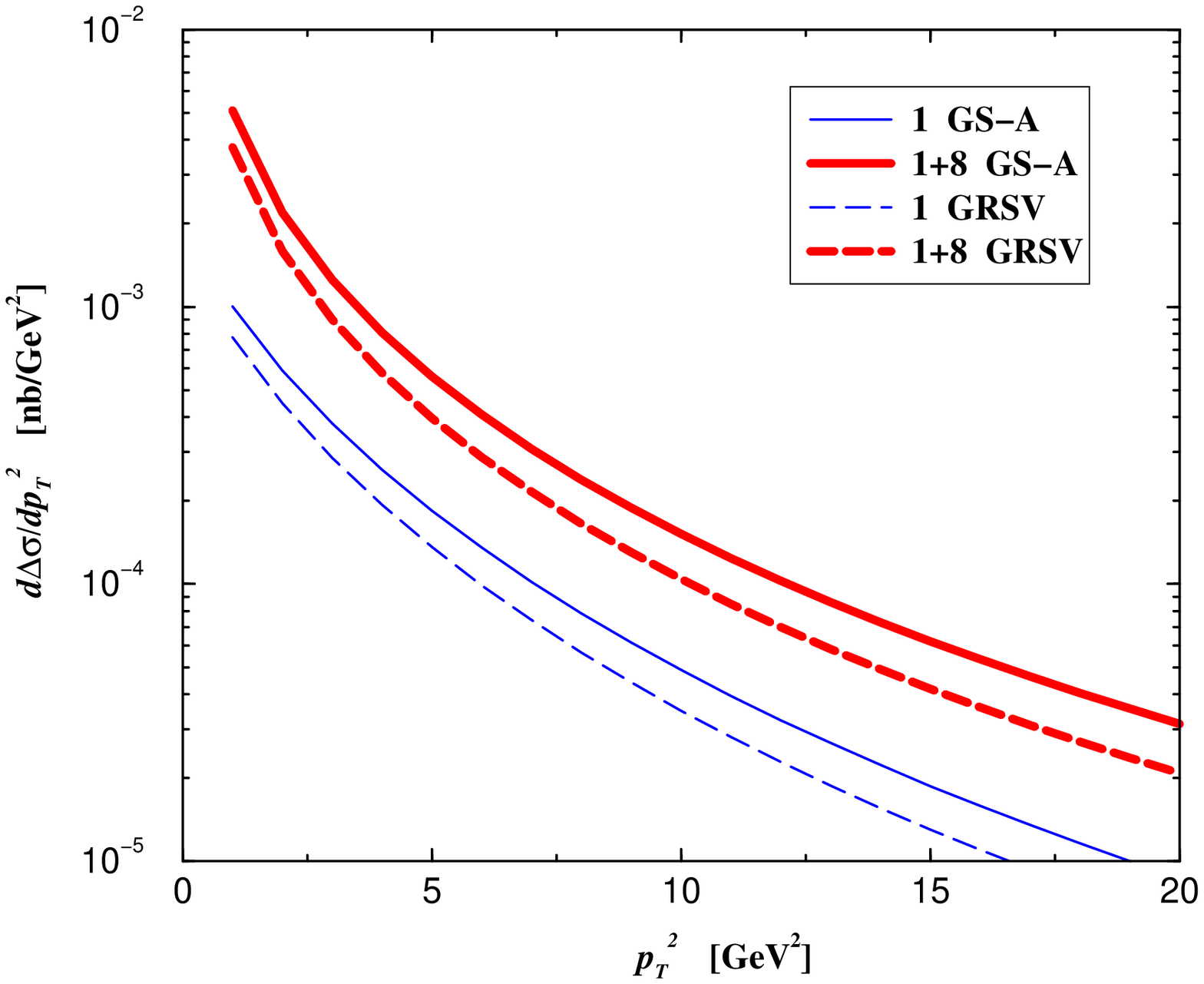}
    \epsfxsize=7cm
    \epsfbox{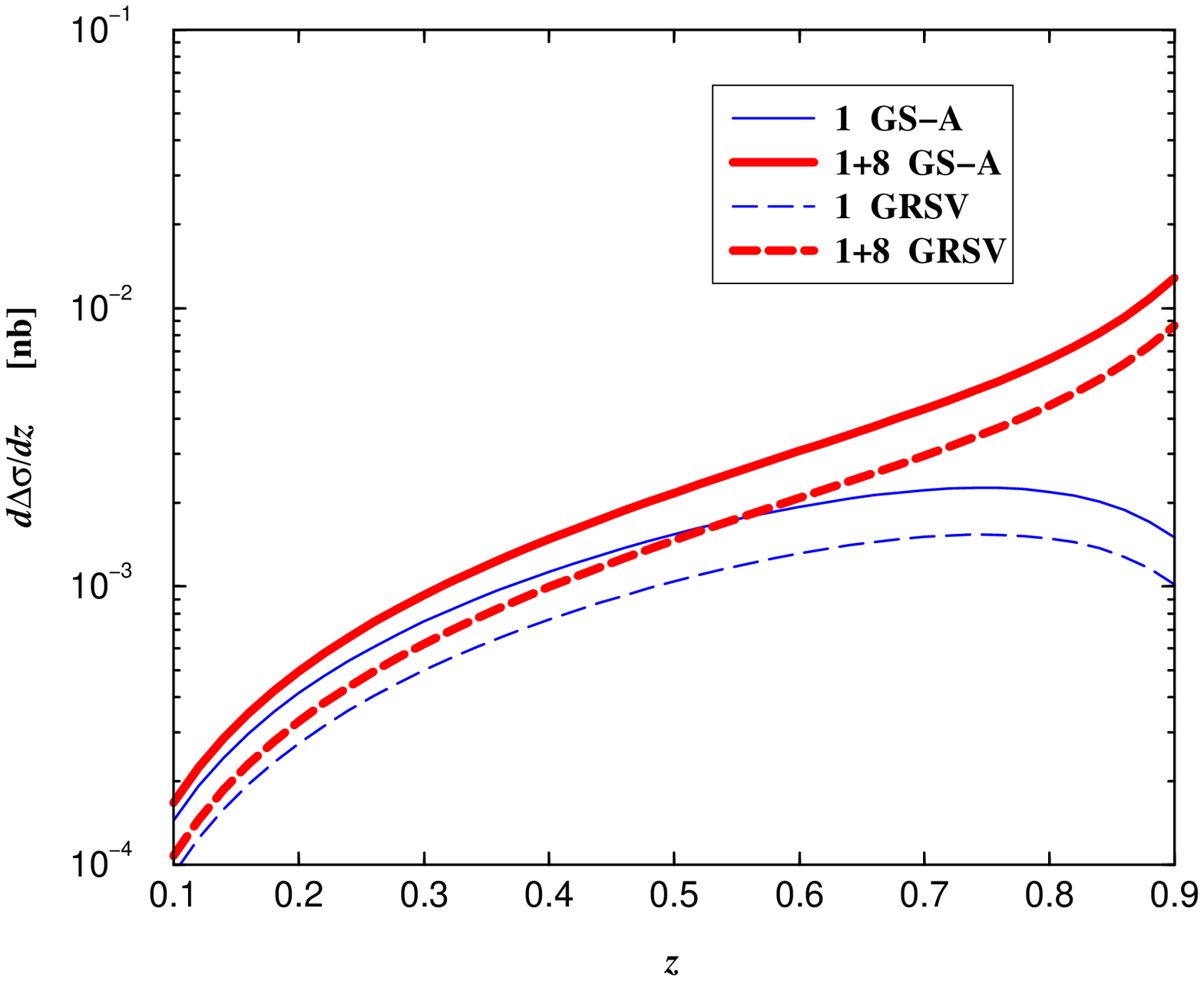}
\end{center}
\vspace{-0.4cm}
\caption[junk]{The spin--dependent differential cross section 
as a function of $p_T^2$ (left) and $z$ (right). 
The bold lines represent the sum of the color--singlet and octet 
contributions, while the normal lines represent the color--singlet 
contribution only. 
The solid and dashed lines show the case of set A of GS \cite{GS96} 
parameterization and the ``standard scenario'' of GRSV \cite{GRSV96} 
parameterization, respectively. 
\label{fig:pol_pt_ep}}
\end{figure}

Setting $Q^2 =4m_c^2$ with a charm quark mass $m_c =1.5$ GeV and the 
relevant center--of--mass energy $\sqrt{s}=300$ GeV, we calculated the
spin--independent and spin--dependent cross sections and double 
spin asymmetry. 
The spin--dependent cross sections are shown in 
Fig. \ref{fig:pol_pt_ep} as a function of $p_T^2$ (left panel), and as 
a function of $z$ (right panel), where $z$ is defined as 
$z=p_{{}_{P}}\cdot p_{\psi^{\prime}}/p_{{}_{P}}\cdot q_{\gamma}$ with 
$p_{\psi^{\prime}}$ being the momentum of outgoing $\psi^{\prime}$. 
In this calculation, we have required the kinematical cut on $p_T$ as 
$p_T^2 > 1$ GeV for $z$ distribution in order to suppress the 
contributions from the diffractive process and higher twist contribution. 
We used the GS set A \cite{GS96} and 
GRSV ``standard scenario'' \cite{GRSV96} parameterizations for the 
polarized gluon distribution function, and the GRV parameterization 
\cite{GRV95} for the unpolarized one. 
In these figures, the solid and dashed lines show the case of set A of 
GS and the ``standard scenario'' of GRSV, respectively. 
Bold lines represent the sum of the color--singlet and octet 
contributions, while the normal lines represent the color--singlet 
contribution only. 
%
\noindent

As shown in Fig. \ref{fig:pol_pt_ep}, the dominant contribution comes 
from the color-octet $^1S_0$ state. 
We can see that the sum of the color--singlet and --octet contribution is 
larger than the color--singlet one alone by a factor of about 5 in the whole 
$p_T^2$ regions. 
It is remarkable that the difference of the cross section due to these 
two production mechanisms is larger than the one of the 
parameterization models for the polarized gluon distribution functions. 
Therefore, this polarized reaction could be another independent test 
of the NRQCD factorization approach. 
In addition, as shown in Fig. \ref{fig:pol_pt_ep} (left panel), it might 
be effective even for testing the model of polarized gluon 
distributions, if we have precise values of the NRQCD matrix elements 
and can observe $p_T^2$ distribution of the cross sections precisely. 

The $z$ distribution of the spin--dependent cross section is also shown 
in Fig. \ref{fig:pol_pt_ep} (right panel). 
Here we do not see big difference between the color--singlet and --octet 
contribution in lower $z$ regions. 
Taking account of the ambiguity of the polarized gluon 
density, we cannot distinguish between the color--singlet and --octet 
contribution in these regions. 
In the larger $z$ regions the color--octet contribution rapidly 
increases and we could see a rather big difference for two production 
mechanisms in these regions. 
However, as is well known, the inelastic photoproduction in the NRQCD 
framework has an important issue on kinematical enhancement in the 
region $z \ra 1$ arising from higher order terms in the NRQCD velocity 
expansion \cite{Beneke97}. 
It is considered that the soft gluon resummation becomes important in 
the larger $z$ regions. 
The prediction for the cross sections by the NRQCD factorization approach 
in this regions should be modified with higher order corrections. 
The prediction for the leading order calculation might be not 
reliable for this region. 
Based on these considerations, the $z$ distribution of the 
spin--dependent cross section is not so effective for testing the 
color--octet model. 
As shown in Fig. \ref{fig:pol_pt_ep} (right panel), it is also not 
good for testing the polarized gluon distribution in the proton. 
%
\noindent
\begin{figure}[t]
\begin{center}
    \epsfxsize=7cm
    \epsfbox{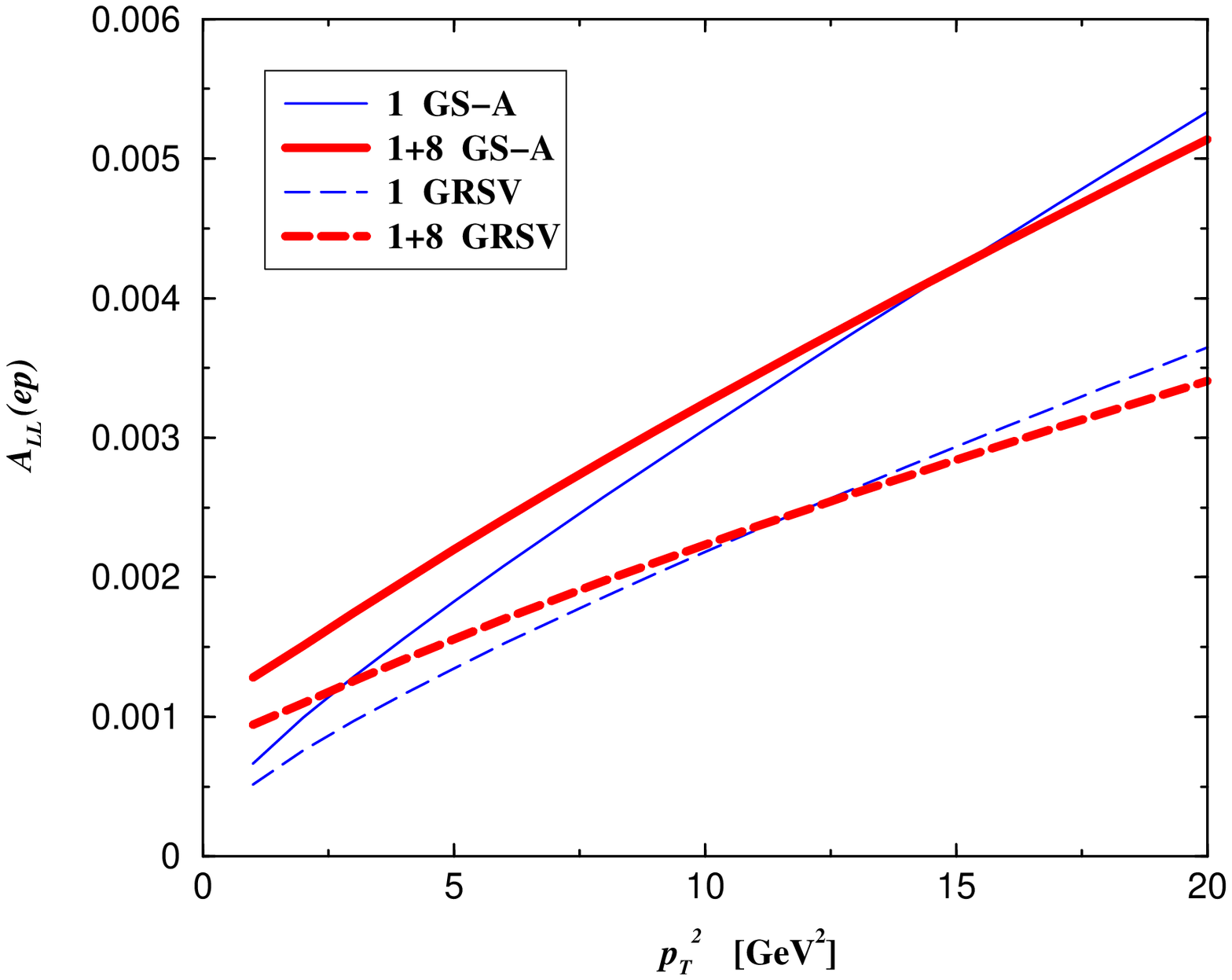}
    \epsfxsize=7cm
    \epsfbox{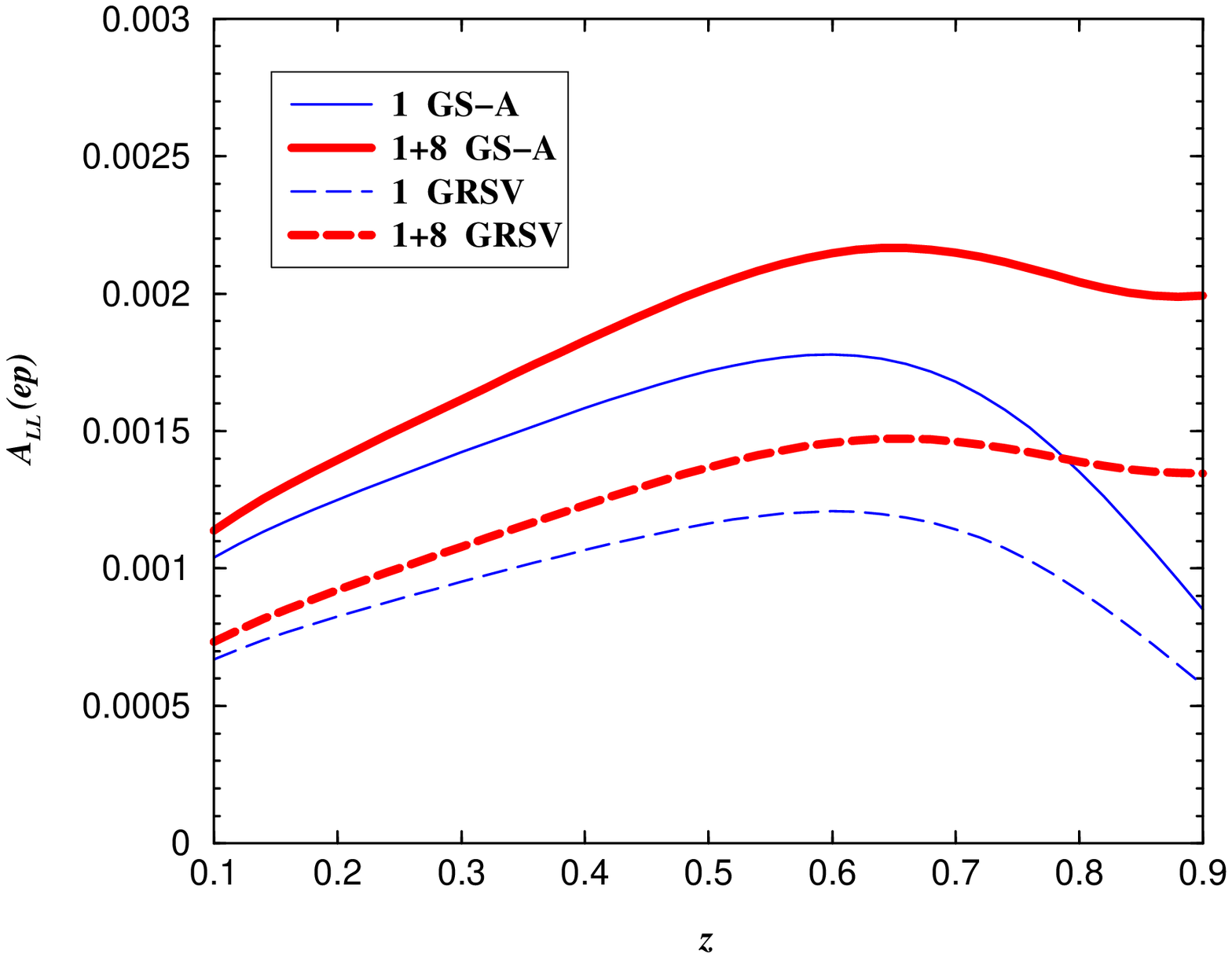}
\end{center}
\vspace{-0.4cm}
\caption[junk]{The double spin asymmetry as a function of $p_T^2$ 
(left panel) and $z$ (right panel). 
Various lines represent the same as in Fig. \ref{fig:pol_pt_ep}. 
\label{fig:all_pt_ep}}
\end{figure}

Next we move to the analysis on the spin correlation for inelastic 
$\psi^{\prime}$ production in polarized $ep$ collisions. 
The double spin asymmetries $A_{LL}$ at $\sqrt{s} = 300$ GeV are 
calculated and presented in Fig. \ref{fig:all_pt_ep} as a function of 
$p_T^2$ (left panel) and $z$ (right panel). 
As seen in the $p_T^2$ distribution of $A_{LL}$, upper two solid lines 
show the calculations for the case of GS set A parameterization. 
The lower two dashed lines represent the ones for the case 
of GRSV ``standard scenario'' parameterization. 
We found that the dependence of the production mechanisms on the $p_T^2$ 
distribution of $A_{LL}$ is quite small in the whole $p_T^2$ regions. 
Instead, the difference due to the models of the polarized gluon density 
is large. 
It becomes larger and larger with $p_T^2$. 
Therefore, we can rather clearly test the models of the polarized gluon 
distribution functions $\Delta g(x)$ by measuring the double spin 
asymmetry for large $p_T$ regions. 
In the $z$ distribution, the difference due to the polarized gluon 
density is again larger than the one due to the production mechanisms, 
though the color--octet contribution dominates over the color--singlet 
one in the whole region of $z$. 

A similar analysis was done by Yuan {\em et al.} for the case of 
$J/\psi$ productions \cite{Yuan00}. 
They have discussed the color--octet contribution to $J/\psi$ 
production and insisted that the process is effective not only for the 
test of the color--octet mechanism but also for the measurement of 
the gluon polarization in a proton. 
However, only direct $J/\psi$ production is considered there, and 
they do not take account of the feedback processes like 
$\psi^{\prime} \ra J/\psi + X$ and $\chi \ra J/\psi + X$, which have 
been shown to play an important role in total $J/\psi$ production 
from experiment. 
For $\psi^{\prime}$ case, there is no such complication, almost all 
contribution come from direct production. 
The advantage of our analysis is that the analysis for $\psi^{\prime}$ 
productions is much clearer than the one for $J/\psi$. 

In summary, in order to obtain information about the polarized gluon 
distribution in a proton, the electroproduction of $\psi^{\prime}$ 
was studied in polarized electron and polarized proton collisions 
by taking account of color--octet contribution. 
As shown in the left panel in Fig. \ref{fig:pol_pt_ep}, 
precise measurement of $d\Delta\sigma / dp_T^2$ for this reaction might 
give another test for the NRQCD factorization approach, if the polarized 
gluon distribution function is sufficiently established. 
Furthermore, if we have precise values of the NRQCD matrix elements, 
we could even test the model of the polarized gluon distribution 
$\Delta g(x)$ in a proton from it. 
On the other hand, $d\Delta\sigma /dz$ is not so effective for testing 
the color--octet model and also $\Delta g(x)$. 
As shown in Fig. \ref{fig:all_pt_ep} (left panel), the $p_T^2$ 
distribution of double spin asymmetry $A_{LL}(p_T^2)$ does not show a 
large difference of the production mechanism with color--octet and 
color--singlet. 
Rather, we see a big difference among the models of the 
polarized gluon distribution function $\Delta g(x)$. 
The $z$ distribution of $A_{LL}$ can be also a good test of 
$\Delta g(x)$, if values of the NRQCD matrix elements are well 
determined. 
Therefore, the process is quite effective for extracting information 
about $\Delta g(x)$. 

\section*{Acknowledgments}

One of the authors (H.W.H.) would like to thank the Monbusho's 
Grand--in--Aid for the JSPS postdoctoral fellow for financial support.

\end{document}